\documentstyle[pra,aps,epsf,twocolumn]{revtex}
\begin{document}
\draft
\font\Bbb =msbm10  scaled \magstephalf
\def\Real{{\hbox{\Bbb R}}}
\def\Complex{{\hbox {\Bbb C}}}
\def\id{{\hbox{\Bbb I}}}

\title{Dynamics of quantum entanglement}

\author{
Karol \.Zyczkowski $^{1,2}$ \cite{poczta4},
Pawe\l{} Horodecki$^{3,}$ \cite{poczta2}, Micha\l{} Horodecki$^{4,}$
\cite{poczta1}, Ryszard Horodecki $^{4,}$ \cite{poczta3}}
\address{$^1$ Centrum Fizyki Teoretycznej PAN,
Al. Lotnik{\'o}w 32/46, 02-668 Warszawa, Poland \\
$^2$ Instytut Fizyki im. Smoluchowskiego,
Uniwersytet Jagiello{\'n}ski, ul. Reymonta 4,
30-059 Krak{\'o}w, Poland \\
$^3$ Wydzia\l{} Fizyki Technicznej i Matematyki Stosowanej,
Politechnika Gda\'nska, 80--952 Gda\'nsk, Poland \\
$^4$ Instytut Fizyki Teoretycznej i Astrofizyki,
Uniwersytet  Gda\'nski, 80--952 Gda\'nsk, Poland}

\maketitle

\begin{abstract}
A model of discrete dynamics of entanglement of bipartite
quantum state is considered. It involves
a global unitary dynamics of the system
and periodic actions of local bistochastic or decaying channel.
For initially pure states the decay of entanglement is accompanied
with an increase of von Neumann entropy of the system.
We observe and discuss revivals of entanglement due to
unitary interaction of subsystems.
For some mixed states having different marginal entropies
of the subsystems we find an asymmetry in speed of entanglement decay.
The entanglement of these states decreases faster, if
the depolarizing channel acts on the
"classical" subsystem, characterized by smaller marginal entropy.
\end{abstract}

\setcounter{figure}{1}

\section{Introduction}
Quantum entanglement is one of the most subtle
and intriguing phenomena in nature \cite{EPR,Sch}.
Its potential usefulness has been demonstrated in
various applications like quantum teleportation,
quantum cryptography and quantum dense coding.
On the other hand, quantum entanglement is a fragile feature,
which can be destroyed by interaction with the environment.
This effect due to  {\it decoherence}
\cite{Zurek}, is the main obstacle for practical implementation
of quantum computing.
A model allowing to study the dynamics of entanglement
in presence of interaction with the environment 
has been recently analysed by Yi and Sun \cite{YS00}.

In this paper we investigate destruction of the entanglement in
a proposed model of discrete dynamics.
We consider a simple bipartite system consisting
of two spin-$\frac{1}{2}$ particles. Only one of them is subjected
to periodic actions of a quantum channel,
which represents the interaction with environment.
As the initial states we choose random states
taken from the ensemble of pure separable
states and from the ensemble of maximally entangled pure states.
We also investigate the time evolution of mixed states
having some special
property. The corresponding system is composed of two subsystems
exhibiting different properties with respect to some
entropy inequality which is satisfied by all classical systems.
One of the subsystems satisfies the
inequality and may be considered "classical", while the other, "quantum"
subsystem violates the inequality.
We investigate an asymmetry in the process of destruction
of entanglement with respect to the subsystem interacting with
the environment.
We demonstrate a possible presence of
revivals of entanglement caused by the global unitary
evolution entangling the subsystems between consecutive actions of
the environment.

The paper is organised as follows. In section II we describe a simple
model of discrete time evolution. In section III we derive bounds on the
entropy increase under the action of the environment.
Then in section IV we analyse the decrease of entanglement versus
increase of the degree of mixing of the initially pure states.
The asymmetry in the entanglement decay depending on
the subsystem subjecting to influence of environment
is described in section V.
In section VI we consider the entanglement revivals.
The results of the paper are discussed in section VII.

\section{Models of time evolution}

In this paper we consider the bipartite state subjected sequential
interactions with environment. They are modelled
by quantum channels, defined as
completely positive linear maps,
preserving the trace of the state \cite{channel}.

Let $\sigma$ be a density operator acting on a
finite-dimensional Hilbert space ${\cal H}$.
The most general form of the quantum channel is
the following transformation $\sigma \rightarrow \sigma'$:
\begin{equation}
\sigma'\equiv \Lambda(\sigma)= \sum_{i=1}^{K} V_i \sigma V_i^\dagger,
\quad {\rm where} \quad  \sum_{i=1}^{K}  V_i^\dagger V_i = \id.
\label{ogolnie}
\end{equation}
If in addition $\sum_{i=1}^{K}  V_i V_i^\dagger= \id$
holds then the channel is called {\it bistochastic}.

Bistochastic channels can be alternatively
defined as channels which do not decrease
von Neumann entropy of quantum states.

A particular example of the bistochastic
channel is given by {\it random external fields}
\cite{Alicki}. They can be written as
\begin{equation}
\sigma' \equiv \Lambda_{R}(\sigma)= \sum_{i=1}^K
p_i A_i \sigma  A_i^\dagger,
\label{ref}
\end{equation}
where $A_{i}$, $i=1,2,\dots,K$ are {\it unitary} operators
and  the vector of probabilities
$\vec{p}=[p_1,\dots,p_K]$ is normalised
\begin{equation}
\sum_{i=1}^{K}p_{i}=1, \quad \ p_{i} \ge 0.
\label{param}
\end{equation}
Such random systems can be described in the formalism of
{\sl quantum iterated function systems} \cite{LZ00}.
The so called Kraus form (\ref{ogolnie})
can be reproduced setting $V_{i}=\sqrt{p_i} A_i$.
It is worth to note that in the case of the most elementary
quantum system described on the Hilbert space
${\cal H} ={\cal C}^2$
the channel is bistochastic {\it if and only if} it
is a random external field (\ref{ref}) (see \cite{Badziag}).
Note that an unitary evolution of the system can be considered as the
simplest case of the bistochastic quantum channel with $K=1$.

There exist, however, many quantum channels which are not bistochastic.
We shall consider
 the following {\it decaying} channel, sometimes called
\cite{Preskill} the {\it amplitude damping channel})
\begin{equation}
\sigma'\equiv\Lambda_D(\sigma)=M_1 \sigma M_1 + M_2 \sigma M_2
\label{decay}
\end{equation}
where the matrices $M_1=
\left[ \begin{array}{cc}
         1 \ \phantom{xx} 0 \\
         0 \phantom{xx} \  \sqrt{p}
       \end{array}
      \right ] $
and
       $M_2=
\left[ \begin{array}{cc}
         0 \ \sqrt{1-p} \\
         0 \phantom{xx}\ \phantom{xx} 0
       \end{array}
      \right ] $
       are written in the standard basis.

Let $\varrho$ denote a mixed state
of a $2 \otimes 2$ system i. e. the density operator
defined on the Hilbert space  ${\cal H}={\cal H}_{A} \otimes {\cal
H}_{B}= {\cal C}^{2} \otimes {\cal C}^{2}$.
The system consists of two subsystems $A$ and $B$
which can represent spin-$\frac{1}{2}$ particles or
two-level atoms.
In our model the unitary dynamics is interrupted by periodic actions of
the environment as shown schematically in Fig. 1.

Discrete time evolution of the state $\varrho$
reads in our model
\begin{equation}
\varrho(n+1)=U \varrho'(n) U^{\dagger}=
U (\hat{\Lambda}(\varrho(n)) U^{\dagger}
\label{dyna}
\end{equation}
where $\hat{\Lambda}=I \otimes \Lambda$ and
the channel $\Lambda$ is either bistochastic (\ref{ref}) or
decaying (\ref{decay}).
Here $U=e^{i \alpha \tilde{H}}$ represents a unitary
transformation which involves an interaction between
both subsystems $A$ and $B$ described by the
 Hamiltonian $\tilde H$.
We use the dimensionless units and
$\alpha$ stands for a coupling parameter.
Subsequently we shall consider the cases with $\tilde{H}$
equal either to $\sigma_{x} \otimes \sigma_{y} \equiv H$
or to $\tilde{H}=\sigma_{y} \otimes \sigma_{x} \equiv H'$.

In general we shall use four types of dynamics
 defined by four different operators $\Lambda$-s in
the formula (\ref{dyna}).
Three of them will be random external fields $\Lambda_R$ (\ref{ref}),
all defined by the same set of $K=4$
unitary operators: $A_{1}=\id $, $A_{2}=\sigma_{1}$,
$A_{3}=\sigma_{2}$,
$A_{4}=\sigma_{3}$ (where $\sigma_{i}$ denote Pauli
matrices), but with different vectors of probability:
(\ref{param}):
\begin{eqnarray}
&&\vec{p}^{(1)}=[1-\epsilon,0,0,\epsilon] \nonumber \\
&&\vec{p}^{(2)}=[1-\epsilon,0,\frac{\epsilon}{2},\frac{\epsilon}{2}]
\nonumber \\
&&\vec{p}^{(3)}=[1-\epsilon,\frac{\epsilon}{3},\frac{\epsilon}{3},
\frac{\epsilon}{3}], \
0\leq \epsilon \leq 1.
\end{eqnarray}

Each dynamics depends on
two continuous parameters:
$\alpha$ contained in $U=e^{i\alpha {\tilde H}}$ governing the
unitary dynamics and
$\epsilon$, included in the vector of probabilities,
and describing the strength of the coupling with the environment.
Additional discrete index
$j$ labels the different vectors of probability,
$\vec{p}^{(j)}$.
For these three models of dynamics we
shall use the compact notation
$\Theta_{\alpha,\epsilon}^{j}$.
The fourth dynamics denoted by $\Theta_{\alpha,p}$
is defined by putting in formula  $(\ref{dyna})$
the decaying channel (\ref{decay}).
Dynamics involving  the operation $U$ with
``reflected'' (i. e. obtained from
$H$ by permutating subsystems) Hamiltonian
$H'= \sigma_{y} \otimes \sigma_{x}$
will be denoted by the same symbols with
only one change: $\Theta \rightarrow \tilde{\Theta}$.

{\it Remark.-}
If $\alpha$ is equal to zero,
then the unitary operation $U$ in (\ref{dyna})
is reduced to identity transformation.
In particular, it can be seen that the dynamics $\Theta_{0, \epsilon}^{3}$
corresponds to periodic action of {\it depolarizing channel}
\cite{Bennett}.

Now the essence of our study is the following:
we consider composite quantum systems subjected to the
local interaction with the environment, which
acts on one subsystem only.
We investigate, how the decay of the entanglement in the
system  depends on the initial state and the type of the dynamics.
In particular we analyse, to which extent the decrease of
the mean entanglement is reflected
by the evolution of von Neumann entropy of the system.

\section{Bounds on entropy increase under local channel}

We start establishing bounds for the increase of
von Neumann entropy.

{\it Proposition .-} Under a local action of the quantum channel
 $\varrho_{AB} \rightarrow (\id \otimes
\Lambda) \varrho_{AB}$, the increase of
the von Neumann entropy $\Delta S$ for
 a bipartite $n \otimes m$ state is bounded by
\begin{equation}
\Delta S \equiv S(\varrho_{AB}^{out}) -  S(\varrho_{AB}^{in})\leq
S(\varrho_{A}^{in}) -  S(\varrho_{AB}^{in}) +\log m,
\label{b1}
\end{equation}
where $S(\varrho_A)$ denotes the entropy of the subsystem $A$.
In particular, if the system is separable then
$\Delta S \leq \log m$.

{\it Proof.-}
By the definition the local channel is trace preserving, hence it does not
change the density matrix of the first subsystem.
Thus  $\varrho_{A}^{out}= \varrho_{A}^{in}$
and the same holds for the corresponding entropies.
Then from subadditivity of the entropy we have
\begin{eqnarray}
S(\varrho_{AB}^{out})\leq S(\varrho_A^{out}) + S(\varrho_B^{out})=
S(\varrho_A^{in}) + S(\varrho_B^{out})\leq \nonumber \\
S(\varrho_A^{in}) + S^{B}_{max}\equiv S(\varrho_A^{in}) + \log m.
\end{eqnarray}
We get the first inequality in the Proposition by subtracting
$S(\varrho_{AB}^{in})$ from both sides of the above inequality.
For separable states one always has
$S(\varrho_{AB}^{in}) -  S(\varrho_{A}^{in}) \geq 0$
\cite{redund} which simplifies
(\ref{b1}) to $\Delta S \leq \log m$
as expected.

Note that a sequence of quantum channels acting locally forms a
quantum channel acting locally too.
So the proposition works also for the
dynamics $\Theta_{0,\epsilon}^{i}$ and $\Theta_{0,p}$.
Moreover, from Eq. (\ref{b1}) we see
that the entropy of initially pure separable state
$\varrho_{AB}^{in}$ cannot exceed $\log m$.

\section{Entanglement versus degree of mixing}

In this section we study the time evolution of entanglement
and compare it with the time evolution of von Neumann entropy.
To characterise the degree of entanglement
we use the {\sl entanglement of formation}
introduced by Bennett et al. \cite{Bennett}.
For any $2 \otimes 2$ mixed state this quantity may be computed
analytically as shown by Hill and Wootters \cite{Wootters}.
In this case the entanglement
of formation $E$ (or shorter, the entanglement) varies from zero
(separable states) to $\ln 2$ (maximally entangled states),
so in the figures we used the rescaled variable $E/\ln 2$.

Our results were obtained
by averaging over ensembles of random initial states.
They were generated according to natural measures on:

(i)  $6$ dimensional manifold of all pure states for $2 \otimes 2 $
problem,

(ii)  $3$ dimensional manifold of maximally entangled pure states,

(iii) $4$ dimensional manifold of separable pure states.

Numerical experiments have shown that the samples of $100$
initial
states, generated  randomly as described in the appendix, were
sufficient to receive reliable results.

\subsection{Bistochastic channels}

As shown in \cite{volume,vol2}
the mean entanglement of mixed states
decreases monotonically with
increasing degree of mixing.
Due to interaction with the environment the initially pure states
become mixed: their von Neumann entropy, $S(\varrho)=-{\rm
Tr}(\varrho\ln\varrho)$, grows in time.
Thus it is natural expect a corresponding monotonous decay
of the mean entanglement. This  indeed takes place,
as shown in Fig. 2 in absence of the unitary
dynamics, $(\alpha=0)$.
Initial states were taken randomly from the entire space of pure states,
so in accordance with \cite{vol2}, the initial mean entanglement is close
to $(\ln 2)/2$. The parallel processes of decay of the entanglement and
increase of the entropy are accelerated, if the
parameter $\epsilon$
describing the interaction with environment increases.

\vskip -1.4cm
\begin{figure}
\hspace*{-1.6cm}
\vspace*{-6.9cm}
\epsfxsize=10.0cm
\epsfbox{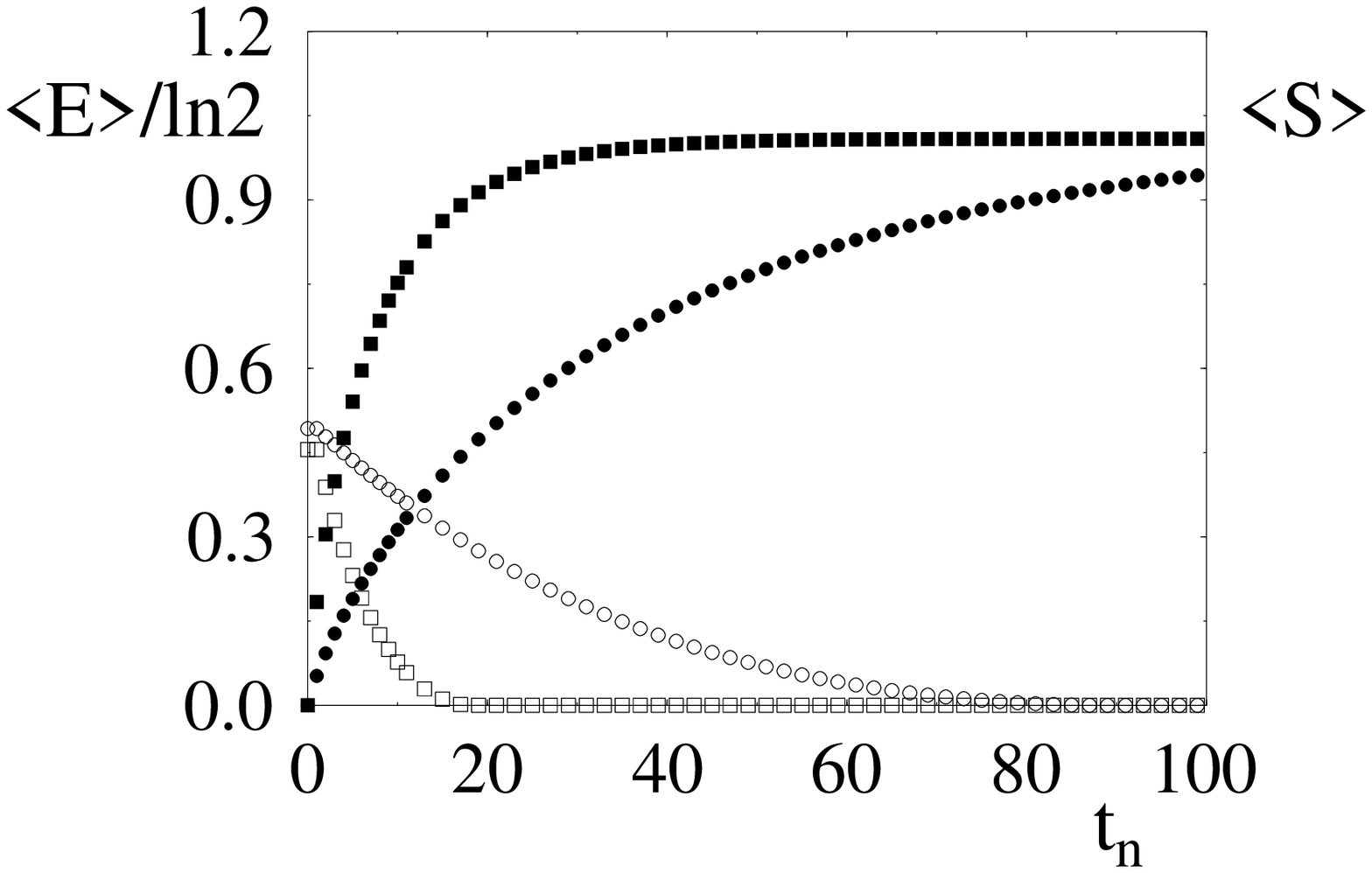}
\caption{Dynamics of quantum entanglement for system
$\Theta^{2}_{0,\epsilon}$.
   Mean entanglement of formation $\langle E\rangle $ (open symbols)
and von Neumann entropy $\langle S \rangle $ (closed symbols)
  averaged over a sample of $100$ random pure states shown as functions
of  discrete time $t_n$.
No unitary evolution is present, ($\alpha=0$).
Parameter $\varepsilon$,  controlling the
interaction with environment
is  set to $0.01 (\circ)$ or $0.05 (\Box)$.}
\label{fig2}
\end{figure}

For initially  maximally entangled pure states (case (ii))
a similar dependence is represented by circles in Fig. 3.
Here $\langle E(0)\rangle=\ln 2$.
The picture changes when unitary evolution is involved.
The latter  leads to oscillations of entanglement
of formation,  reflected in the time  evolution
of entropy.
 The frequency of oscillations is proportional to
$\alpha$. The larger this parameter,
the faster the unitary evolution $U$ rotates
the states $\varrho$ from and into the convex set of separable states.
In the case of entropy, oscillations are only due to changes
of the second derivative i.e. entropy is still monotonically decreasing.
This is not the case for entanglement $E$,
which  can also be seen in Fig. 4
for several individual initial states (without averaging).
For short times the curve for $\alpha=0$
(no unitary evolution) seems to constitute an
envelope for all other curves.

\vskip -2.5cm
\begin{figure}
\hspace*{-2.5cm}
\vspace*{-0.2cm}
\epsfxsize=10.0cm
\epsfbox{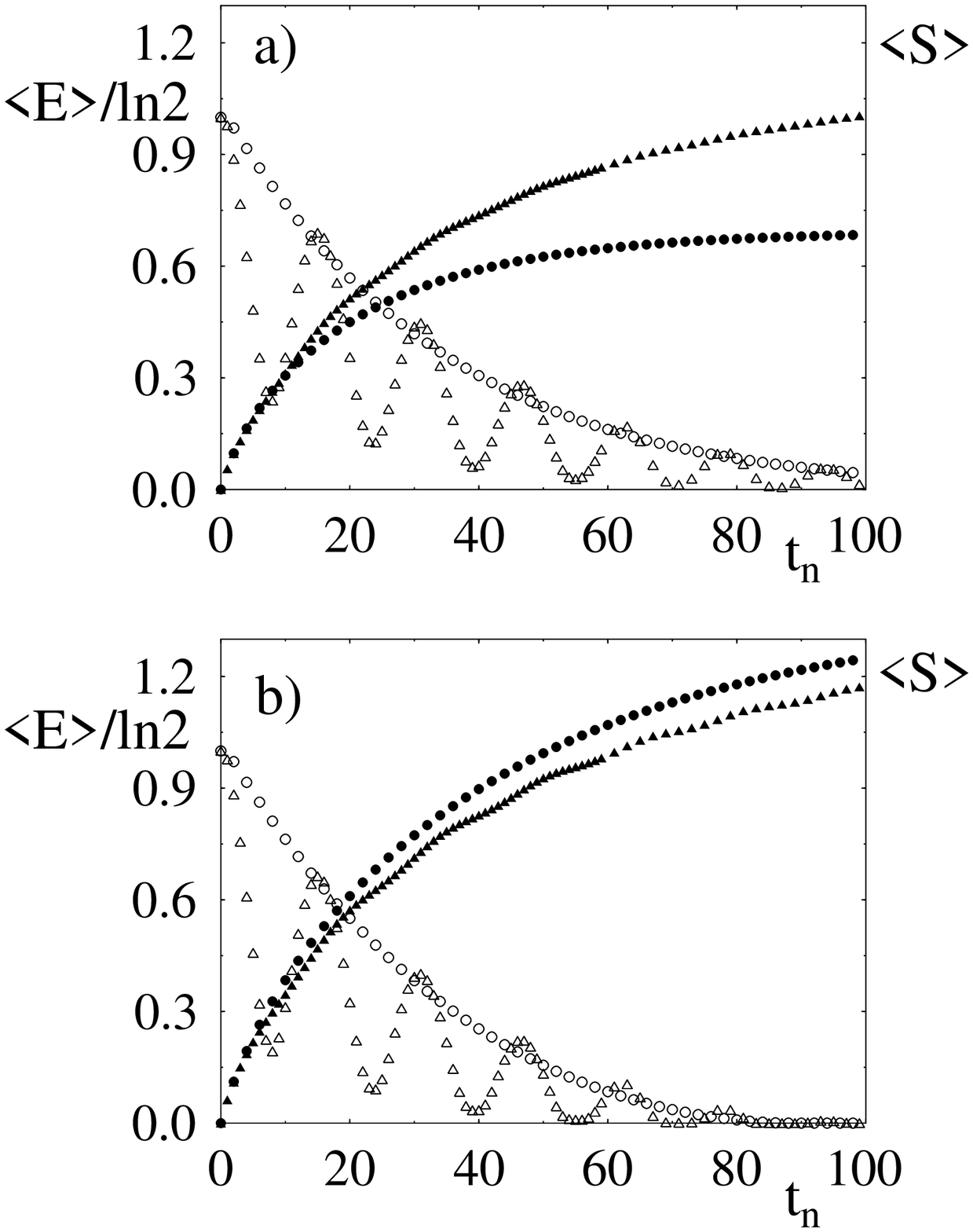}
\caption[Dynamic3]{As in Fig. 2 for a sample of $100$
maximally entangled states ($E(0)=\ln(2)$) with $\varepsilon=0.01$;
$\alpha=0.0(\circ)$ and $\alpha=0.1(\triangle)$
for channels described by a) $\vec{p}^{(1)}$ and b) $\vec{p}^{(2)}$.
Observe how the influence of the unitary dynamics depends on the kind
of the channel.}
\label{fig3}
\end{figure}
\medskip

It is worth to emphasise a significant difference between
$\Theta_{\alpha,\epsilon}^{1}$
(Fig. 3.a) and $\Theta_{\alpha,\epsilon}^{2}$ (Fig. 3.b).
In the former case the presence of unitary evolution
can accelerate the process of entropy increase. In the latter,
on the contrary, switching on unitary evolution results
in slower increase of the mean entropy.
The iteration of the channel
$\Theta_{\alpha,\epsilon}^{1}$
preserves both the number and the position
of the nonzero component in $\vec{p}$. It is not the case for
$\Theta_{\alpha,\epsilon}^{2}$, for which
two Pauli matrices generate
the entire algebra of unitary matrices $A_{i}$ involved.

\vskip -2.1cm
\begin{figure}
\hspace*{-1.5cm}
\vspace*{-5.2cm}
\epsfxsize=10.0cm
\epsfbox{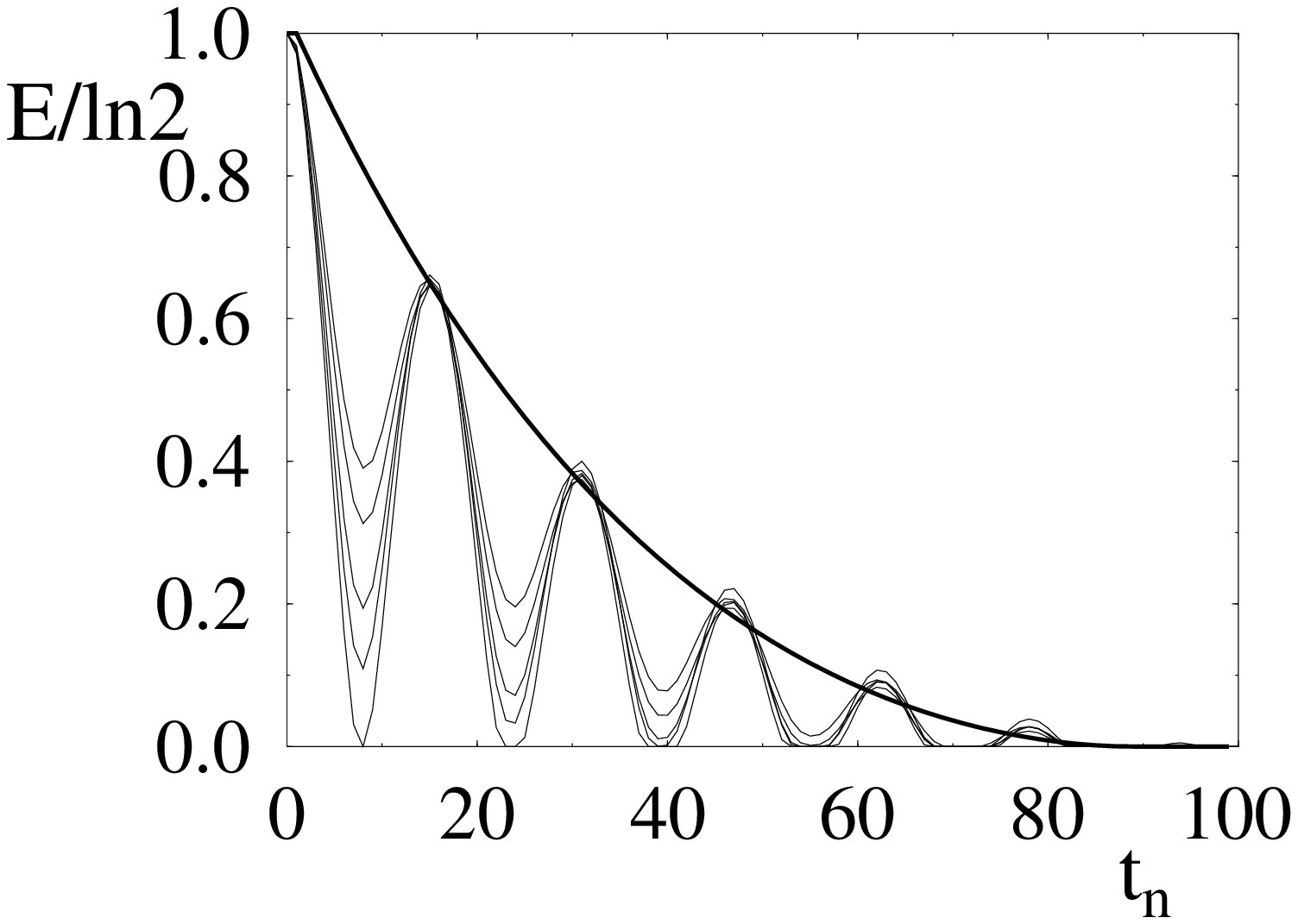}
\caption[Dynamic4]{Dependence of entanglement of formation on
time for several randomly chosen maximally entangled pure states.
The unitary dynamics $U=\exp(i\alpha \tilde{H})$
is governed by the parameter $\alpha$.
Here $\varepsilon=0.01$ in $\vec{p}^{(2)}$, and  $\alpha=0.1$ (narrow
lines). Reference bold line represents the case of no unitary dynamics
$(\alpha=0)$, for which the dynamics of entanglement does not depend on
the initial state.}
\end{figure}
\medskip

Consider now the case (iii), of initially separable states,
presented in Fig. 5.
The presence of the unitary evolution may increase the mean
entanglement, initially equal to zero.
However, there is one difference more;
for {\it both} dynamics $\Theta_{\alpha,\epsilon}^{1}$ (Fig. 5.a)
and  $\Theta_{\alpha,\epsilon}^{2}$ (Fig. 5.b)
presence of the unitary dynamics
accelerates the process of increase of entropy.
In absence of the unitary dynamics ($\alpha =0$) the entropy does
not exceed the value
 $\ln 2$, in accordance to our proposition proved in section III.

Obtained results show that
the oscillations of the mean entanglement $E$ are
anti-correlated with the oscillations of the entropy $S$.
It was also checked that if $\alpha$ is kept constant,
 but the Hamiltonian is chosen randomly then the
oscillations of entanglement are smeared out.
It means that effects of quantum coherence are destroyed
 and the destruction of entanglement occurs faster.

\vskip -8.1cm
\begin{figure}
\hspace*{-2.3cm}
\vspace*{-1.0cm}
\epsfxsize=10.0cm
\epsfbox{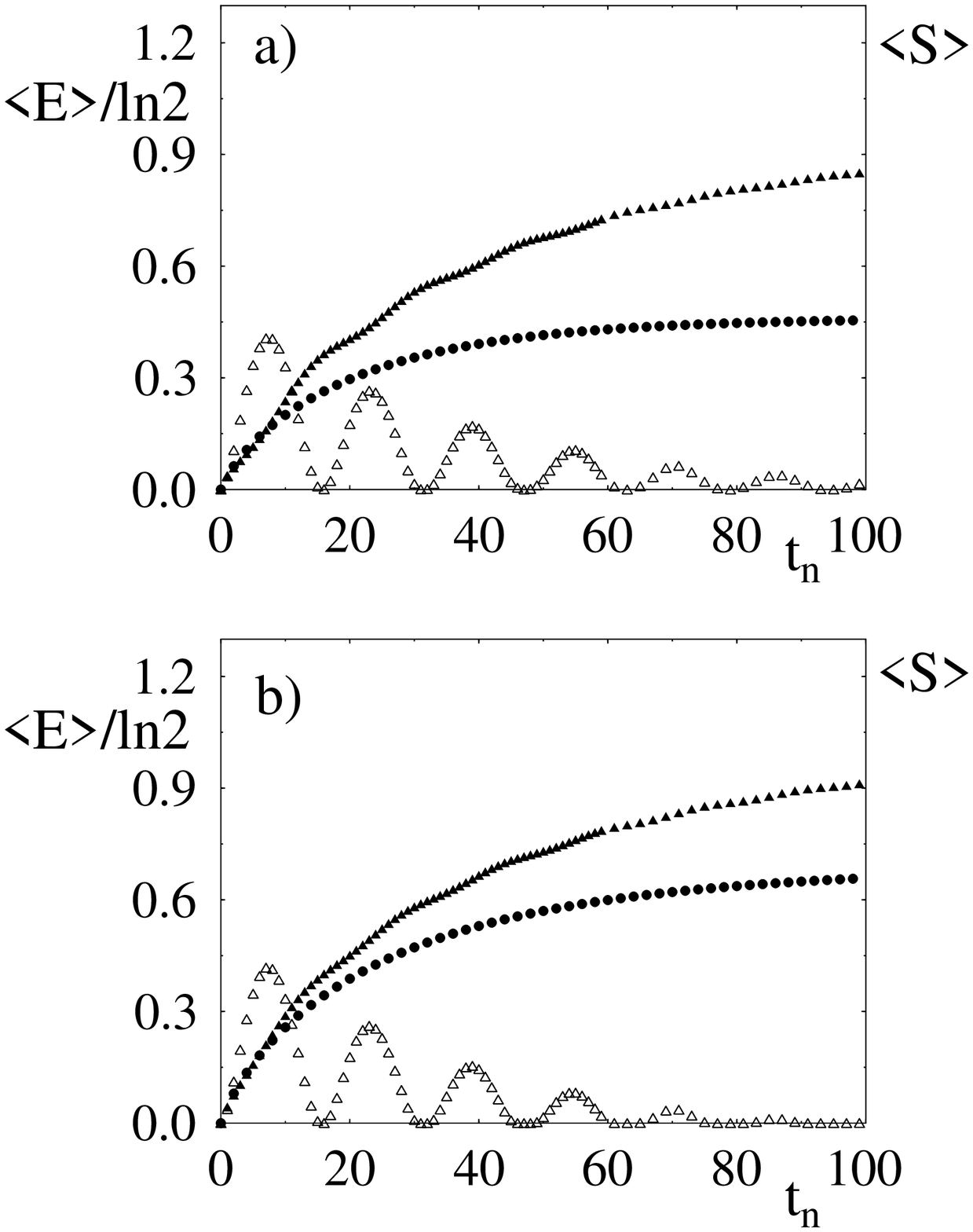}
\caption[Dynamic5]{
As in Fig. 3 for a sample of $100$ initially separable pure
states ($E(0)=0$). In absence of unitary dynamics,
($\alpha=0$), the entanglement equals to zero.}
\label{fig5}
\end{figure}
\medskip

\subsection{Decaying channel}

Figure 6 presents results obtained for
the amplitude damping channel  (\ref{decay}).
In the absence of the unitary evolution
($\alpha=0$) the mean entropy, $\langle S \rangle$,
averaged over the entire manifold of pure states (case (i)),
does not tend monotonically to its maximal value.
At $t_n\sim 20$
the entropy reaches its maximum and then
decreases to its limiting value about $0.3$
 (see full circles in Fig. 6b).
This is due to the fact that for the {\it decaying} channel the
entropy of the system may decrease.

Numerical data received by averaging over the set of
maximally mixed states (case (ii), diamonds) and
the set of separable pure states (case (iii), squares)
are shown in Fig. 6a.
Observe that the steady state limiting values of the von Neumann entropy,
$\langle S \rangle$,  represents the initial average entanglement
$\langle E \rangle$.
Indeed, in absence of the unitary evolution the perturbed subsystem is
eventually dumped to the ground state. So finally 
the state of the system is a product one consisting of the ground state
of the affected subsystem and  the reduced density matrix
of the unperturbed subsystem.
Thus, after the averaging procedure, one gets the averaged von Neumann entropy
of  the subsystem not subjected to action of the channel.

A random choice of initially pure states of the composite system induces
a certain measure in the space of the reduced density matrices 
\cite{braun}.
As proved recently by Hall \cite{Ha} 
 the natural rotationally invariant
measure on the space of $N=4$ pure states induces
a uniform measure in the {\sl Bloch ball} representing the
density matrices for $N=2$.
Denoting the spectrum of reduced matrices by $\{ 1/2-r,1/2+r \}$
we may write more formally, $P(r)=24r^2$ for $r\in [0,1/2]$.
The von Neumann entropy, averaged over this measure
equals $1/3$ \cite{Ha},
  in agreement with the numerical data presented in Fig.6.

\vskip -1.3cm
\begin{figure}
\hspace*{-1.9cm}
\vspace*{-0.9cm}
\epsfxsize=10.0cm
\epsfbox{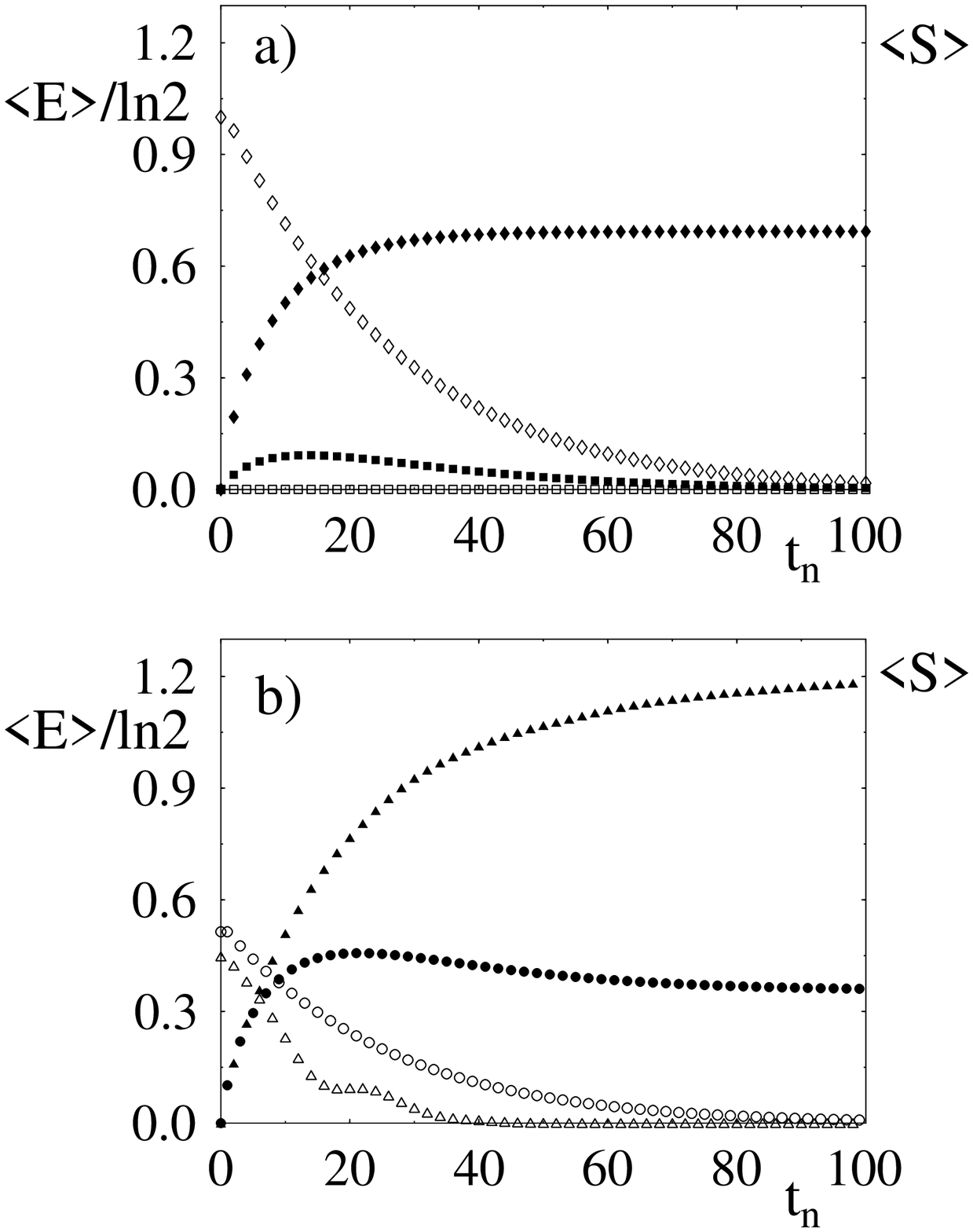}
\caption[Dynamic6]{
As in Fig. 2 for samples of $100$ pure states
subjected to the Kraus channel $(4)$
with $p=0.05$ and $\alpha=0.0$. Initial states 
drown randomly from the ensembles of
a) $(\Box )$ --  separable pure states; a) $(\diamond )$ --
maximally entangled pure states, and  b) $(\circ)$ --
ensemble of all pure states.
Case (i) with unitary evolution,
$\alpha=0.1$, is denoted by $(\triangle)$
in panel b).
}
\end{figure}
\medskip

For non-zero values of $\alpha$ we observe the
oscillations of the mean entanglement, caused by the unitary evolution.
It is interesting, however, that the presence of unitary evolution allows
the final entropy to be maximal (see full triangles in Fig. 6b).
It means that the presence of the decay channel is
{\it completely masked} by the unitary
interaction between both subsystems.

\section{Asymmetry of entanglement decay}

We shall consider here dynamics of mixed states having  an
intriguing property.
Namely we choose a quantum bipartite system, which
{\it violates some entropy inequality only with respect
to one of both subsystems}.
Let us recall first that
 the information gain resulting from the measurement
of any of subsystems of a quantum state with classical correlations
is not
greater than the gain obtain form measurement performed on the entire system.
This classical feature is characteristic of quantum separable
states. They do satisfy the following two inequalities concerning von
Neumann entropy \cite{redund,alfa}:
\begin{equation}
 S(\varrho_{AB})\geq S(\varrho_{A}),
\label{sntr}
\end{equation}
and
\begin{equation}
 S(\varrho_{AB})\geq S(\varrho_{B}),
 \label{sntr2}
\end{equation}
where $\varrho_{A}$ and $\varrho_{B}$ denote the reduced density
matrices, e.g. $\varrho_{A}\equiv Tr_{B}(\varrho_{AB})$.
Now we shall focus on the following family of states introduced
in \cite{Tata}.
They can be written
as $\varrho^{(1)}:=q|\Psi_{1}\rangle\langle\Psi_1|
+ (1-q)|\Psi_{2}\rangle\langle\Psi_2|$, $0<q<1$,
with normalised pure state vectors $|\Psi_{1}\rangle=a|00\rangle+
\sqrt{1-a^{2}}|11\rangle$
and $|\Psi_{2}\rangle=a|10\rangle+\sqrt{1-a^{2}}|01\rangle$
with $0<a<1$.
In the standard basis, ($|00\rangle$, $|01\rangle$, $|10\rangle$,
$|11\rangle$), the corresponding density matrix takes the form

\scriptsize
\begin{equation}
\varrho^{(1)}=\left[ \begin{array}{cccc}
        qa^2  &   0 &
\  0   & \ qa\sqrt{1-a^2} \\
    0 & \  (1-q)(1-a^2) & \
(1-q)a\sqrt{1-a^2} \ & 0 \\
       0 &  (1-q)a\sqrt{1-a^2}  & (1-q)a^2
&  0 \\
     qa\sqrt{1-a^2} &0  &   0  & q(1-a^2)
\end{array}       \right ]. \label{stan} \end{equation}
\normalsize

Let us take $a^2>q>\frac{1}{2}$. Then the first inequality (\ref{sntr}) is
violated, while the second one (\ref{sntr2}) is not. Thus the composite
system can be called 'quantum'
with respect to the subsystem $A$ and 'classical' with respect to the subsystem
$B$. One may  then expect
 that the bipartite system will loose entanglement
in different ways, depending on whether the environment interacts with
classical or quantum subsystem. Intuitively one could guess that the
entanglement should be more robust if the noise affects the classical
subsystem.

Here we studied the system $\rho^{(1)}$ for
$q=3/5$ and $a^2=3/4$. Then von Neumann entropy of the entire system,
 $S=s(2/5)\approx 0.673$ is greater than the entropy of the
classical subsystem $B$ for which $S_B=s(1/4)\approx 0.562$,
and smaller than the entropy of the quantum subsystem,
 $S_A=s(9/20)\approx 0.688$,
 where $s$ stands for the binary Shannon entropy,
 $s(x):=-x\ln x -(1-x)\ln(1-x)$.
 We analysed the time evolution of this quantum system
in presence of a depolarizing channel $\Theta _{3,\epsilon}^0$ given by
(\ref{ref}). In the theory of error correcting codes it
 is one of the most popular models of environment induced noise.
The evolution of entanglement for the state $\varrho^{(1)}$
is represented by stars in Fig.7. In this case the bistochastic channel
$\tilde{ \Lambda}$ acts on the 'classical' subsystem $B$.
To investigate a possible asymmetry of the entanglement decay
we consider the state $\rho^{(2)}$, for which
both subsystems are exchanged. More precisely, all elements of both density
matrices are equal, apart from
$\rho^{(2)}_{23}=\rho^{(1)}_{32}$ and $\rho^{(2)}_{32}=\rho^{(1)}_{23}$.
The corresponding dynamics of $\rho^{(2)}$
is denoted by crosses in Fig. 7. In this case the noise interacts with the
'quantum' subsystem $A$. The magnification in the inset
reveals the asymmetry in the time evolution.
Observe that
the attack on the 'classical'  part of the system
is more harmful to the entanglement properties of the system.
This counter intuitive effect, called subsequently
 {\it anomalous entanglement decay} (AED),
links quantum and classical
features of the state from information-theoretical point of view.

\vskip -1.8cm
\begin{figure}
\hspace*{-1.5cm}
\vspace*{-6.6cm}
\epsfxsize=9.5cm
\epsfbox{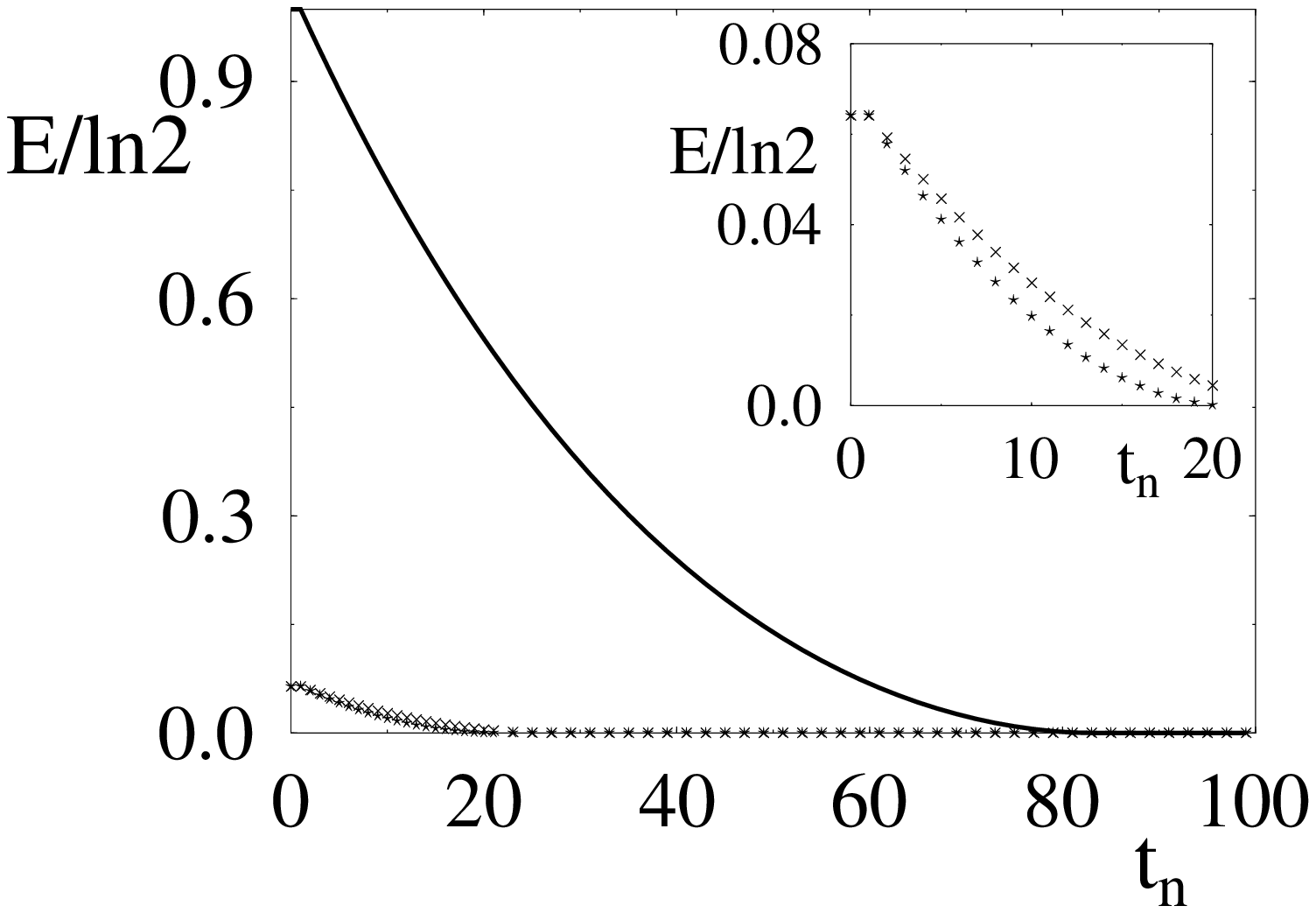}
\caption[dynamic7]{Comparison of the dependence of the entanglement of
formation for the state $\varrho^{(1)}$
with $a^2=3/4$; $q=3/5$
$(\star)$ and $\varrho^{(2)}$ $(\times)$.
The bistochastic channel  $\vec{p}_{3}$ with $\varepsilon=0.01$
interacts with the 'classical' subsystem $B$ in the former case,
and with the 'quantum' subsystem $A$ in the latter case.
 Solid line represents the behaviour of a maximally
entangled state $\rho_{\rm max}$. Magnification
of the initial dependence provided in the inset
reveals the asymmetry of the entanglement decay.
}
\end{figure}
\medskip

Let us recall that any $2 \otimes 2$ system may be described
by two Bloch vectors, representing locally
both subsystems, and a correlation matrix $T$, which
represents the projection of the composite
system onto the family of mixtures of
maximally entangled states (see \cite{pra}).
A possible explanation of AED
should take into account the fact that the local action
of environment
changes both the Bloch vectors, the correlation matrix,
as well as their relationship. A depolarizing  channel may
affect in a similar way both local parameters,
but it may distinguish, (in sense of
the destruction of the entanglement),
the correlation parameters with respect to the side of the action.

It should be noted that, regardless which part is subjected to
the noise, the entanglement of mixed states $\rho^{({1})}$ and
$\rho^{(2)}$ decreases slower
than the entanglement of the maximally entangled states
(bold line in Fig. 7).
This is due to the fact that the latter decreases
fast for short times and slow at longer time scales,
for which the initially pure state gets mixed.
It is thus  instructive to compare the shape
of the bold line starting from $t_N \approx 60$
with the symbols representing the initial decay of entanglement of
the states  $\rho^{(i)}$.

\section{Amplifying the processes: entanglement revivals.}

Consider now the depolarizing dynamics
 $\Theta _{\alpha,\epsilon}^3$ with
an unitary operation involved, ($\alpha \neq 0$),
affecting either subsystems $A$ or $B$.
To compare the dynamics of both symmetric mixed states
$\varrho^{(1)}$ and $\varrho^{(2)}$
we study their unitary interaction governed by
the Hamiltonians: $H=\sigma_{x} \otimes
\sigma_{y}$, and the reflected one
$H'=\sigma_{y} \otimes \sigma_{x}$.

\vskip -1.1cm
\begin{figure}
\hspace*{-1.9cm}
\vspace*{-0.9cm}
\epsfxsize=9.5cm
\epsfbox{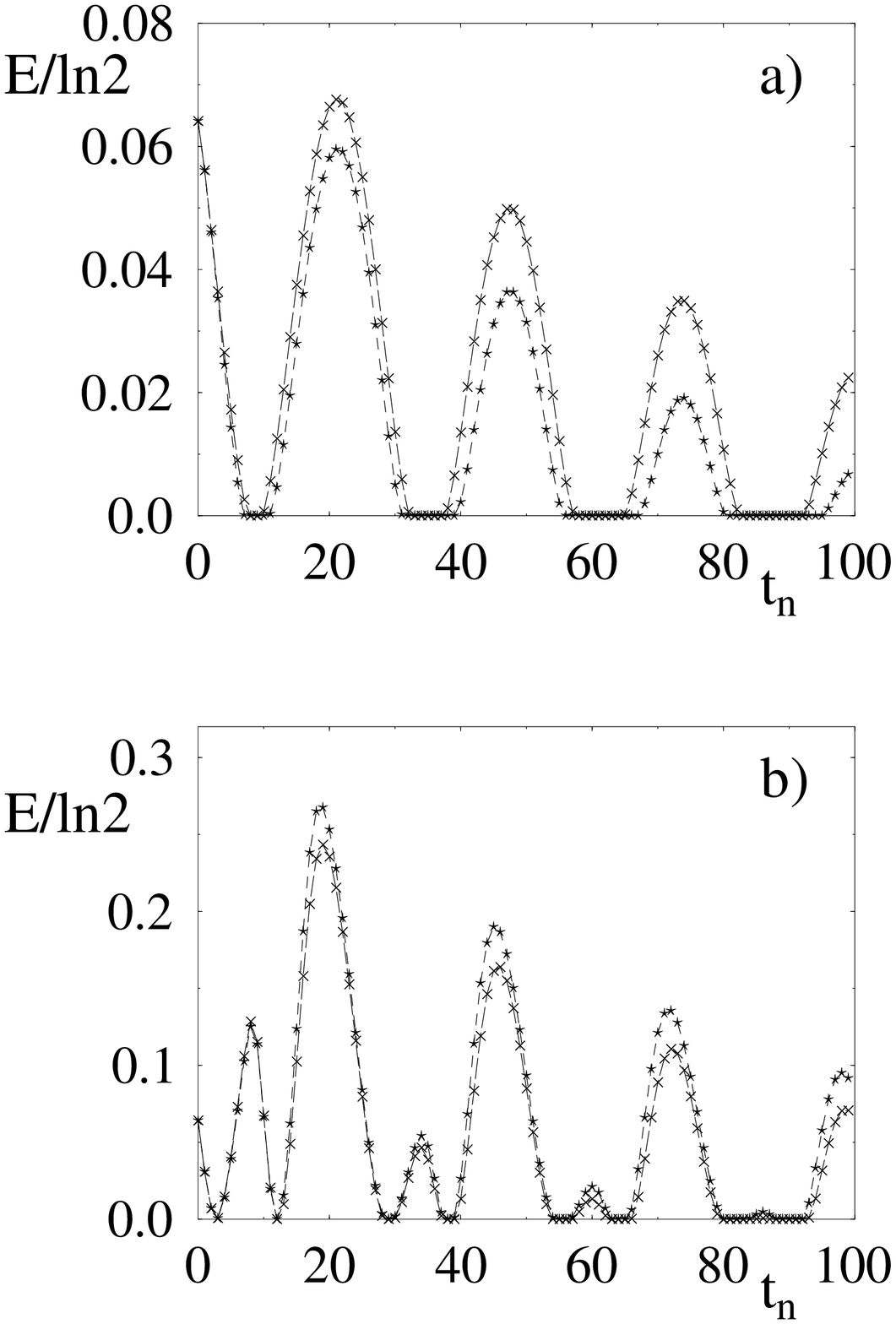}
\caption[Dynamic8]{
Unitary dynamics and asymmetry
of entanglement decay:
a) the state $\varrho^{(1)}$ defined by parameters $a^2=3/4$ and $q=3/5$
subjected to the bistochastic channel $\vec{p}_{3}$ with $\varepsilon=0.002$
and unitary dynamics $H$ with $\alpha=0.06$ ($\star$);
the symmetric state $\varrho^{(2)}$
interacting with the reflected Hamiltonian $H'$ ($\times$).
Panel b) shows the data for reflected unitary dynamics; the Hamiltonians
$H'$ and $H$ are exchanged.}
\label{fig8}
\end{figure}

Let us consider two cases:

(a) the noise parameter $\epsilon$ is much less than the parameter $\alpha$
characterising the unitary interaction,

(b) both parameters are of the same order of magnitude.

Numerical results obtained in the weak noise case (a)
are presented in Fig. 8. and 9.
The revivals of the entanglement,
caused by the unitary interaction,
 are  manifestly visible, since
 the strength of the interaction with the
environment $\epsilon=0.002$ is much less then the
parameter  $\alpha=0.06$ governing the unitary dynamics.
Note  the characteristic  entanglement ${\it plateaus}$, if
the analysed state travels across the set of the separable states
and the entanglement attains its minimal value equal to zero.
The  effect of anomalous entanglement decay
is clearly visible in Fig. 8.a,
where the entanglement decays faster
if the environment interacts with the classical subsystem.
This contrasts the situation shown in Fig. 8.b,
for which the unitary evolution
is due to the reflected Hamiltonian $H'$
and the exposure of the 'quantum' subsystem
to the action of the environment action is
more damaging for the entanglement.

\vskip -1.1cm
\begin{figure}
\hspace*{-1.9cm}
\vspace*{-0.9cm}
\epsfxsize=9.5cm
\epsfbox{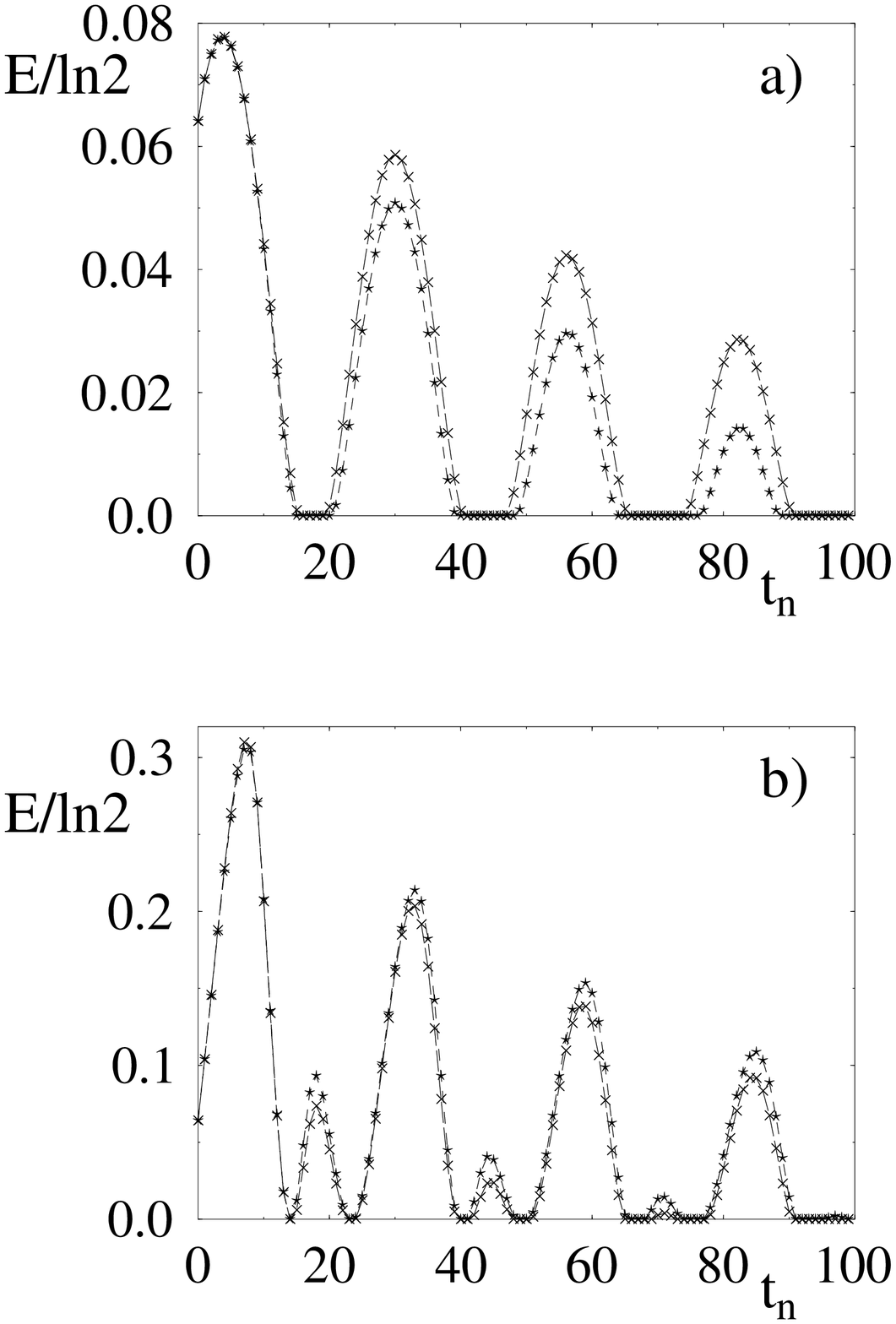}
\caption[Dynamic11]{
As in Fig. 8 for
$\varepsilon=0.002$ and $\alpha=-0.06$,
i. e. the process runs back in time.
Observe that maxima in Fig. 8.a correspond to minima in
Fig. 9.a and vice versa.}
\label{fig9}
\end{figure}

It is instructive to analyse the same system with the
unitary evolution reversed in time. Such a case,
 obtained by a change of the parameter $\alpha \rightarrow -\alpha$,
is presented in Fig. 9.
The general character of the evolution is kept. The
significant difference is that here the entanglement is {\it amplified}
at the very beginning which may have practical
consequences if we are interested in short times of the process.
Note that the figures 8.a. and
 9.a reflected along the vertical line at $t_n=0$
  (respectively,  8.b and reflected 9.b)
exhibit some kind of symmetry
with respect to the initial moment.

What happens if we allow the strength of
 the coupling with the environment
to be comparable with the parameter of the unitary
interaction? This situation, corresponding to
 the case (b), is illustrated
in Fig. 10. Here some interesting qualitative changes occur.
The AED effect is present in the case shown in Fig. 10a;
at the beginning the entanglement
disappears faster when the `classical'
part of the system is affected by the environment.
 Moreover, in this case the
entanglement disappears completely and never revives.
If  the `quantum' subsystem interacts with the environment,
a single {\it entanglement revival} occurs.

\vskip -1.1cm
\begin{figure}
\hspace*{-1.9cm}
\vspace*{-0.9cm}
\epsfxsize=9.5cm
\epsfbox{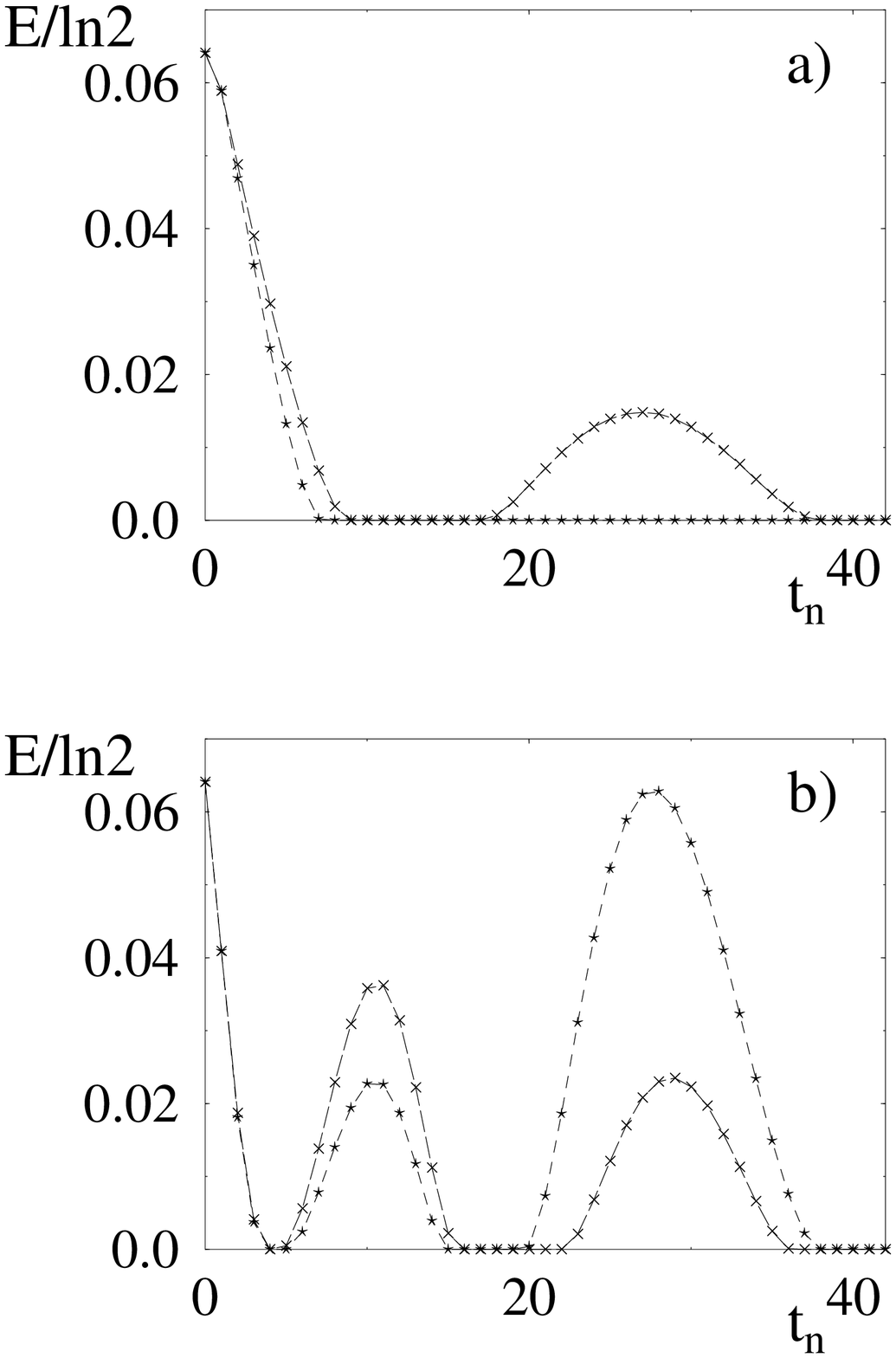}
\caption{As in Fig. 8 for
$\varepsilon=0.01$ and $\alpha=0.04$.
}
\label{fig10}
\end{figure}

In the complementary case, for which $\varrho^{(1)}$ interacts with
the reflected Hamiltonian $H'$ (see 10.b),
we observe a special kind of competition:
for short times the entanglement is smaller, if the quantum
subsystem is perturbed. For longer times, the roles are interchanged,
and the oscillations of the entanglement
are damped faster, if the classical subsystem interacts with the
environment.

In general one can see that the pictures
corresponding to the cases (a) and (b) are
qualitatively different depending on the
ratio $\epsilon / \alpha$.
This fact may be related  to the observation concerning the
processes of decoherence. Depending on the
relation between two coupling parameters
the so called {\it pointer basis}
is determined either by the internal self-Hamiltonian
of the system or by the Hamiltonian of the
 interaction with environment \cite{Zurek}.

\section{Discussion}

We investigated the behaviour of entanglement of bipartite spin-$\frac{1}{2}$
system  subjected to  periodic action of the environment.
The process of destruction of entanglement of initially pure states
is accompanied by increasing of von Neumann entropy. The asymptotic
value of the entropy depends on the form of the interaction with the
environment. For strongly mixing bistochastic channels,
(e.g. $\Theta^2$ and $\Theta^3$) the entropy achieves the maximal value $\ln
4$.  If the decaying channel is involved, the
entropy gets its maximum and then it  monotonically decays
to the asymptotic value, which reveals the initial entanglement of the system.

If the internal unitary evolution entangling the system is present,
the  decay of the entropy due to the decaying channel
can be replaced  by the process of mixing the state more and more.
The general feature of the time evolution is
that the entanglement decreases as the system becomes more mixed.
This corresponds to the results recently obtained in \cite{volume,vol2},
where it was shown that the mean entanglement of quantum states,
averaged over a sample of
mixed states with the same von Neumann entropy,
decreases with the degree of mixing.
The presence of the internal unitary evolution leads to the
revivals of the entanglement and to
suppression (or acceleration) of the entanglement decay.

Perhaps the most intriguing is the character of asymmetry
of the time evolution of the entanglement. For some initial mixed
states consisting of two non-equivalent subsystems,
the entanglement decays faster, if the environment
interacts with the 'classical' subsystem,
which satisfies the entropic inequality.
Many years ago Schr{\"o}dinger
considered entanglement of pure state as a property of
having  both subsystems less informative for the observer,
than the composite system. Mixed states (\ref{stan})
considered here exhibit this property only with respect to
one subsystem \cite{Tata}.
Our results show  that the action of environment
to the `classical' subsystem is sometimes
more harmful to the entanglement. In this case one can thus say
that the {\it quantum entanglement
runs away faster through the classical door}.

In the context of the above discussion some general
questions emerge. Consider a quantum entangled state $\varrho$
with, say, $S(\varrho_{A}) > S(\varrho_{B})$,
not necessarily violating the inequality (\ref{sntr}).
Under which conditions the
entanglement is less robust to
action of the environment on this
subsystem, for which the entropy of the 
reduced operator is smaller?
 How is it related to a possible violation of
the von Neumann entropy inequality  by subsystem $A$?
What happens if instead of the inequalities (\ref{sntr},\ref{sntr2})
one applies the generalised $\alpha$-entropies
inequalities (\cite{alfa,Tata,rank}
satisfied for classical systems?
All these questions seem to be important for deeper understanding
of the dynamics of quantum entanglement.

It would be also interesting to analyse the role
of entropic asymmetric states like (\ref{stan})
in context of quantum communication.
In fact these states have only one coherent information
positive \cite{Nielsen,Lloyd} (see also \cite{Shannon}).
For the corresponding quantum channels this
might imply an asymmetry in the transfer of quantum information
with respect to its direction, ($A\to B$ or $B \to A$).

Finally, the obtained results  show that
even the simplest bipartite systems may exhibit
non-trivial properties form the point of view of the information theory.
In this context it would be important to
investigate further the dynamics of mixed entanglement,
in particular, by taking into account the phenomenon of
bound entanglement \cite{bound}.

It is a pleasure to thank the European Science Foundation
and the Newton Institute for a support during our stay
in Cambridge, where this work has been initiated.
K.\.Z is supported by Polish Committee for Scientific
Research, contract No. 2 PO3B 072 19.
M. P. R. H. are supported by Polish Committee for Scientific Research,
contract No. 2 P03B 103 16 and by European Community
under the IST project EQUIP, contract No. IST-1999-11053.

\appendix

\section{Random pure states}

In this appendix we present algorithms allowing one
to generate random quantum states distributed uniformly
at the entire space of pure states, the manifold of
separable pure states and the space of
 maximally entangled pure states. We concentrate here on the
simplest $2 \otimes 2 $ problem, but the algorithms below can be
easily generalised for higher dimensions.

\subsection{Generation of random pure states}

The set of pure states of a $4$ dimensional Hilbert space
forms a complex projective space $\Complex P^{3}$, on which a
natural, unitarily invariant measure exist.
To generate random pure states according to such a measure
on this $6$ dimensional space we take a vector of a random
unitary matrix distributed according to the Haar measure on $U(4)$.
The Hurwitz parametrization \cite{Hu87} gives

\begin{eqnarray}
| \Psi \rangle = (\cos \vartheta_3,
\sin \vartheta_3 \cos \vartheta_2  e^{i \varphi_3}, \nonumber \\
\sin \vartheta_3 \sin \vartheta_2 \cos\vartheta_1  e^{i \varphi_2},
\sin \vartheta_3 \sin\vartheta_2  \sin \vartheta_1 e^{i \varphi_1} ),
\label{param3}
\end{eqnarray}
where $\vartheta_k \in [0,\pi/2],$
and $\varphi_k  \in [0, 2\pi)$ for $k=1,2,3$.

A uniform distribution over almost all of
 $\Complex P^{3}$ is obtained by
choosing the uniform distribution of the 'azimuthal' angles;
$P(\varphi_k)= 1/ 2\pi$.
In the analogy to the volume element on the sphere
 the  'polar' angles $\vartheta_k$
should be taken in a nonuniform way, with the probability density
\cite{Hu87}
\begin{equation}
P(\vartheta_k)=k \sin(2\vartheta_k)
  (\sin \vartheta_k)^{2k-2}
\label{density2}
\end{equation}
for
$\vartheta_k \in [0,\pi/2]$,  $k=1,2,3$.
In practice it is convenient to use auxiliary independent random
variables
$\xi_k$ distributed uniformly in $[0,1]$ and to set
$\vartheta_k={\rm arcsin}\bigl(\xi_k^{1/2k} \bigr)$.
Above formula with $k=1,2,\dots,N-1$ allows one to
get a natural distribution on $\Complex P^{N-1}$ \cite{PZK98}.

\subsection{Random separable pure states}

Any $2 \otimes 2$ pure separable state may be
written as $|\Psi_s\rangle = |\psi_1\rangle \otimes |\psi_2\rangle$,
where $|\psi_1\rangle$ and $| \psi_2\rangle$ are $N=2$, one-particle pure
states.
The $4$ dimensional manifold of separable states has thus a simple
structure of a Cartesian product
$\Complex P^1 \times \Complex P^1$. A uniform measure
on this manifold is obtained be taking both states $|\psi_i\rangle$
distributed uniformly (and independently) at the Bloch sphere,
 $\Complex P^1 \sim S^2$.

Working in the standard basis,
\begin{equation}
 |\Psi_s\rangle= U_1 \otimes U_2 |(1,0,0,0) \rangle,
\label{sepr}
\end{equation}
 where $U_1$ and $U_2$ denote
two independent random unitary matrices distributed uniformly on $SU(2)$.
This parametrisation describes the  entire $4$D manifold
of the separable pure states.

\subsection{Random maximally entangled states}

In an analogous way we may represent the maximally entangled states
as
  \begin{equation}
 | \Psi_e\rangle= \id \otimes U_1
|(0,1,1,0)/\sqrt{2} \rangle.
\label{ent1}
\end{equation}
It is easy to see that for this states the reduced density matrix
is proportional to identity matrix, and the entropy of entanglement
achieves its maximum $\ln 2$. The states obtained by
 a symmetric operations $U_1 \otimes \id$
 are also maximally entangled.
 Using the standard representation of
$U_1$ we parametrise
maximally entangled states by  \cite{KZ00}
\begin{equation}
|\Psi_{e1}\rangle =
\frac{1}{\sqrt{2}}\left[
\begin{array}{l}
\cos\vartheta e^{i\varphi_1} \\
\sin\vartheta e^{i\varphi_2} \\
-\sin\vartheta e^{-i\varphi_2} \\
\cos\vartheta e^{-i\varphi_1}
\end{array}
\right] .  \
\label{ent2}
\end{equation}
The angles $\varphi_i$ are distributed uniformly in
$[0, 2 \pi)$, whereas
according to (\ref{param3})
$P(\vartheta)=\sin(2\vartheta)$
for $\vartheta \in [0,\pi/2]$.
Note that the standard element of the volume on the two
sphere $dS=\sin\theta  d\theta d\varphi$
is written in a rescaled variable $\theta=2\vartheta$.
Given maximally entangled state corresponds to
a single unitary matrix $U_1$ pertaining to  $SU(2)$,
but the $3$-D manifold of the maximally entangled states
has the topology of the real projective space,
$\Real P^3\sim U(3)/U(1)$\cite{VW00}.


\newpage
\setcounter{figure}{0}

\begin{figure}
\hskip0.5cm\begin{picture}(85,170)(0,-170)
\put(45,-20){$\varrho'_{0}$}
\put(55,-53){$\hat\Lambda$}
\put(110,-48){$U$}
\put(45,-87){$\varrho_{0}$}
\put(70,-20){\vector(4,-3){80}}
\put(65,-80){\vector(0,1){60}}
\put(64,-19){$\circ$}
\put(64,-88){$\circ$}
\put(169,-20){$\varrho'_{1}$}
\put(146,-53){$\hat\Lambda$}
\put(201,-48){$U$}
\put(169,-87){$\varrho_{1}$}
\put(161,-20){\vector(4,-3){80}}
\put(156,-80){\vector(0,1){60}}
\put(155,-19){$\circ$}
\put(155,-88){$\circ$}
\put(260,-20){$\varrho'_{2}$}
\put(237,-53){$\hat\Lambda$}
\put(292,-48){$U$}
\put(260,-87){$\varrho_{2}$}
\put(252,-20){\vector(4,-3){80}}
\put(247,-80){\vector(0,1){60}}
\put(246,-19){$\circ$}
\put(246,-88){$\circ$}
\put(351,-20){$\varrho'_{3}$}
\put(328,-53){$\hat\Lambda$}
\put(383,-48){$U$}
\put(351,-87){$\varrho_{3}$}
\put(343,-20){\vector(4,-3){50}}
\put(338,-80){\vector(0,1){60}}
\put(337,-19){$\circ$}
\put(337,-88){$\circ$}
\put(64,-115){\vector(1,0){370}}
\put(64,-115){\line(0,1){6}}
\put(155,-115){\line(0,1){6}}
\put(246,-115){\line(0,1){6}}
\put(337,-115){\line(0,1){6}}
\put(390,-145){$t_n$}
\put(64,-130){$0$}
\put(155,-130){$1$}
\put(246,-130){$2$}
\put(337,-130){$3$}
\end{picture}
\caption[Conclusive teleportation]{Discrete model of periodic
dynamics (\ref{dyna}), (cf. Fig. 8.1, 8.2 of
Ref. \cite{Penrose}).
Interaction with the environment $\hat \Lambda$ transforms the state
 $\varrho_{n}$ into
$\varrho'_{n}$ and then the unitary transformation $U$ maps it into
$\varrho_{n+1}$.
}
\end{figure}

\end{document}